# Efficient Confinement of Ultraviolet Light into the Self-Assembled, Dielectric Colloidal Monolayer on a Flat Aluminum Film

*Seungwoo Lee[1*], Juyoung Kim[2]*

[1]SKKU Advanced Institute of Nanotechnology (SAINT) & School of Chemical Engineering, Sungkyunkwan University (SKKU), Suwon 440-746, Republic of Korea

[2]Department of Neurobiology, Stanford University School of Medicine, Stanford, CA 94305, USA

*Email: seungwoo@skku.edu

**Abstract**

Here we propose the efficient confinement of ultraviolet (UV) light into the plasmonic-photonic crystal hybrid, which can be practically developed by the self-assembly of dielectric colloidal nanosphere monolayer onto a flat aluminum (Al) film. Using a numerical approach, we analyzed modal characteristics of each different resonant mode at the UV wavelengths including surface plasmon polariton (SPP) mode and waveguided (WG) mode and tuned these resonant modes from deep to far UV simply by adjusting the size of dielectric colloidal nanosphere. The calculated quality-factor ($Q$-factor) of such plasmonic-photonic crystal hybrid is at least one order of magnitude higher than that of the existing Al nanostructures (Al nanoparticles, nanodisks, nanovoids, or nanogratings) standing on the dielectric substrate. Also, we systematically studied how the amount of native oxide, which can be generated during the general process for the deposition of Al, can influence on both the SPP and WG modes of such plasmonic-photonic crystal hybrid in order to guide strategies for a realistic experimental fabrication and exploitation of relevant optical responses. We anticipate that the theoretical results in this paper enable a promising step in the enhancement of UV light interaction with the nanophotonic structure in a versatile, but highly efficient way.

## 1. Introduction

In contrast to noble metals (e.g., gold and silver), aluminum (Al), due to its d-band located above its Fermi level, allows us to push the near-infrared (IR)/visible-frequency regime of plasmon to the ultraviolet (UV), so as to open up the emerging applications of UV plasmonics [1-11]. In particular, as the relatively high energy level of UV well corresponds to the various electronic transition levels of organic molecules, the development of advanced molecular spectroscopy based on enhanced fluorescence sensing or surface enhanced Raman scattering (SERS) can be further boosted up by taking advantage of UV wavelength plasmonics. Furthermore, its natural abundance together with high compatibility with CMOS process makes Al highly promising material candidate for a new opportunity of other transformative applications in UV plasmonics as well [1-11]. This unique ability to instantiate UV plasmonics has driven the creation of Al nanostructures and exploitation of the relevant plasmonic modes; there has been little advancement, however, of rationally designed Al nanophotonic structures, capable of high quality-factor ($Q$-factor) at UV wavelength [2−7, 11]. Although the most of Al-mediated UV plasmonics has been indeed exploited mainly by the localized modes in individual Al nanoparticles/nanodisks/nanovoids or the propagating modes in Al gratings, the obtainable $Q$-factor of such general plasmonic structure on a dielectric substrate has been still relatively low (generally less than 50) [2−7, 11]. This relatively low $Q$-factor of UV light resonance, especially, limits the availability of UV plasmon-enhanced molecular spectroscopy or sensing; thus, the currently available $Q$-factor of Al nanophotonic structure, in practice, needs to be further increased.

This paper theoretically proposes the efficient UV light confinement and the resultant high $Q$-factor by using the plasmonic-photonic crystal hybrid. The coupling of dielectric photonic crystal structure to a flat metallic film (i.e., plasmonic-photonic crystal hybrid) has proved to be a versatile platform for the efficient confinement of the light into the nanoscaled space and the achievement of high $Q$-factor (at least over 100) at the near infrared (NIR)/visible wavelength [12−15]. Particularly, the dielectric colloidal monolayer on a flat noble metal film (e.g., Au), supporting both dielectric optical (whispering gallery-like resonant waveguided modes, abbreviated as WG) and surface plasmonic resonance (surface plasmonic polariton modes, abbreviated as SPP), has actually shown to conceive a high $Q$-factor (up to 600) at a NIR or visible wavelength [12−15]. This ability to contain high $Q$-factor together with a versatile-to-implement fabrication method makes the structural motifs, consisting of dielectric colloidal monolayer and a flat metal film, highly amenable to the various practical applications of plasmonic-photonic crystal hybrid including sensing platform, light emitting device, and solar cells [13, 14, 16]. Here, we have extended this strategy to enable the efficient confinement of UV light and the resultant high $Q$-factor with the Al-mediated plasmonic-photonic crystal hybrid, which can be simply developed via the self-assembly of dielectric colloidal monolayer (polystyrene (PS) colloidal nanospheres) on a flat Al film rather than other noble metal films (Figure 1). With respect to the spatial distribution of electric field intensity ($|E|^2$), we confirmed that Al plasmonic-photonic crystal hybrid can support both WG and SPP modes at the UV wavelength. Also, we figured out that $Q$-factor at each WG and SPP modes can be generally as high as several hundreds at the UV wavelengths; the resonant wavelengths of its WG and SPP modes can be easily tuned from deep and far UV merely by adjusting the size of PS nanospheres. Meanwhile, Al can be easily contaminated by oxidization (aluminum oxide, $Al_2O_3$) during the fabrication process (i.e., Al deposition by e-beam evaporation): Al has been generally known to be encapsulated or mixed with oxide according to the deposition condition [11]. Thus, in order to set out the general guide for the experimental fabrication and reproducible exploitation of the relevant optical responses, we systematically analyzed the influence of $Al_2O_3$ to be in Al film (as form of composite) or onto Al surface (as form of encapsulation layer) on the WG and SPP modes of Al plasmonic-photonic crystal hybrid.

**2. Stepwise evolution of optical responses of plasmonic-photonic crystal hybrid**

In order to confirm the possible resonant modes, three different structural primitives were systematically exploited including the hexagonally close-packed, PS colloidal monolayers, which are assembled (i) in air (Figure 1a), (ii) onto 600 μm thick Si wafer (Figure 1b), and (iii) onto 100 nm thick flat Al film/ 600 μm thick Si wafer (Figure 1c). The size of PS colloidal nanosphere was varied from 230 nm to 330 nm (300 nm-sized PS colloidal monolayer was employed as a main structural primitive, herein): 330 nm-sized PS nanosphere was found to be upper limit for conceiving the far UV wavelength resonant modes, as will be discussed later. By using numerical approach (finite-difference time-domain, FDTD), the collective set of optical responses of the plasmonic-photonic crystal hybrid, including the reflection spectra (by Fourier transformation of the time-transient pulse signal), $|E|^2$ spatial distribution, and $Q$-factor ($\omega/\Delta\omega$, where $\omega$ is the resonant frequency and $\Delta\omega$ is the resonant linewidth) [12-15, 17] was systematically rationalized: all of the optical responses was obtained by simulating unit cell (the box highlighted by blue dotted line, as shown in Figure 1d) of the close-packed, PS colloidal monolayer array with the periodic boundary conditions. An ultra-fine grid (at least 60 pts per wavelength in a $x$-, $y$-, $z$-direction) was employed; the time-transient response was collected for a relatively long time (at least 3 ps). The dielectric constants of Si wafer and $Al_2O_3$ were empirically obtained by the ellipsometric measurement, while the PS was assumed to be a non-dispersive dielectric material with $n$ of 1.549. The complex permittivity of Al was characterized by a modified Drude model, $\varepsilon=\varepsilon_\infty - (\omega_p^2/(\omega^2+i\omega\gamma))$, where $\varepsilon_\infty$ is background dielectric constant (high-frequency response), $\omega_p$ is plasma frequency, and $\gamma$ is damping frequency. The adapted Drude model fitting parameters were as follows: $\varepsilon_\infty$ = 3.0 eV, $\omega_p$ = 16.3 eV, and $\gamma$ = 0.8 eV. The employed electric polarization was $x$-axis in Figure 1a; due to the 6-folded symmetric geometry (hexagonally close-packed geometry), such structure does not show the dependency of optical response to two orthogonal polarizations ($x$- or $y$-directions), when the light is normally irradiated ($z$-direction).

Figure 2 summarizes the optical responses of 300 nm-sized PS colloidal monolayers in air (see Figure 2a), including reflection spectra (Figure 2b) and spatial distribution of $|E|^2$ (Figure 2c). In this case, the incident light can be coupled to the individual PS nanospheres by whispering gallery mode [16]. Also, more importantly, the periodically close-packed, PS nanospheres can support both the in-plane waveguided modes (slab guided mode, confined by total internal refraction) and the resonant waveguided modes, as with other 2D photonic crystal slabs [17]. The latter, in turn, results in the coupling of resonant light to the surrounding air (i.e., resonant radiation losses), as evidenced by several peaks (Lorentzian shaped) of reflection spectra (see Figure 2b) [14, 15, 17]. The nature of each resonant radiation mode (marked by WG1, WG2, and WG3 in Figure 2) is revealed with the $|E|^2$ spatial distributions, as presented in Figure 2c; the structural symmetry in the vertical direction makes PS colloidal monolayer to support the purely dielectric resonant modes. Interestingly, the sharp and asymmetric resonant behavior is well observed in the reflection spectra (Figure 2b), owing to the interference between the reflection and the exponentially degraded amplitude of the resonances [17] within the PS colloidal monolayer; the obtainable *Q*-factors vary from 40 to 70 with respect to each resonant radiation mode.

When such PS colloidal monolayer is directly attached to the dielectric Si substrate (Figure 3a), the symmetry of the structure in the vertical direction becomes broken and the relevant efficiency of light confinement is found to be further reduced (see Figure 3b-c). The resonant radiation modes, supported by the PS colloidal monolayer onto a flat Si wafer (Figure 3a), were rationalized by the responses of reflection spectra, presented in Figure 3b (the resonant behaviors are revealed by the deeps rather than peaks of reflection spectra, due to the interference between the resonant radiation modes and the reflected wave from the high refractive index Si wafer); especially, WG1, WG2, and WG3 of these modes show a lowered $|E|^2$ with more oriented spatial distribution toward the Si wafer, compared with a freestanding counterpart (Figure 3b). This is because the resonant waveguided light within PS colloidal monolayer eventually gets coupled to a high refractive index dielectric substrate (Si, *n* of 3.4): this is radiative loss. As a result, the *Q*-factors of such resonant modes tend to be smaller than those of a freestanding PS colloidal monolayer, as evidenced by broaden resonant behavior in reflection spectra. Furthermore, the implementation of a high refractive index dielectric substrate into the PS colloidal monolayer causes the resonant modes to red shift. Meanwhile, such direct coupling of PS colloidal monolayer to the high refractive index substrate newly results in the generation of higher resonant radiation mode, marked by WG4, showing much higher $|E|^2$ and *Q*-factor than other three modes. These four-different modes can be divided into two groups, depending on the $|E|^2$ spatial distribution (Figure 3c): coupled (WG1 and WG3) or non-coupled (WG2 and WG4) modes to the Si wafer substrate.

The resonant behavior of these four-different modes can be further boosted up or reconfigured by the implementation of a flat Al film into the back-plane substrate, as shown in Figure 4a. Figure 4b presents the reflection spectra of the PS colloidal monolayer, assembled onto a 100 nm thick, flat Al layer/Si backing layer; we can clearly observe that the resonant behavior becomes shaper and deeper, indicating enhanced *Q*-factors and coupling efficiencies, respectively. Particularly, the previous WG1 and WG3 can be strongly coupled to the Al substrate layer and transformed into SPP-like resonant modes (re-named as SPP1 and SPP3), as clearly revealed by the $|E|^2$ (upper panel in Figure 4c) and *E*–vector spatial distribution (bottom panel in Figure 4c): the *E*–vectors of both SPP1 and SPP3 modes near the surface of a flat Al film are vertically oriented (transverse magnetic (TM)-like mode), while simultaneously $|E|^2$ is highly concentrated between PS nanosphere and a flat Al film. These further confinements of the UV light into the vicinity between PS nanosphere and Al layer via strongly coupled SPP allows us to obtain the enhanced *Q*-factors (from 48.2 to 126.1 for SPP1; from 32.2 to 66.5 for SPP3). WG2 and WG4 modes, resonant modes formed by the dielectric PS colloidal monolayer, also can be greatly facilitated by their interaction with a flat Al layer. As clearly shown in Figure 4c, in both WG2 and WG4 modes, the $|E|^2$, confined within a dielectric PS spherical nanoparticle, is strongly enhanced, while the relevant *Q*-factors are significantly boosted up, for example, from 46.1 to 475.9 for WG2 and from 183.7 to 590.3 for WG4. These simply conceivable high *Q*-factors merely by the self-assembly of PS colloidal monolayer onto a flat Al film allow us to emphasize one of strengths of such Al plasmonic-photonic crystal hybrid system: these calculated *Q*-factors at the UV wavelengths are at least one order of magnitude higher than those, which are obtainable with the existing Al-based UV plasmonic systems (e.g., Al nanoparticles, nanodisks, nanovoids, or nanogratings) [2−7, 11]. Meanwhile, the implementation of a flat Al layer into the PS colloidal monolayer makes the system to have additional higher modes, indicated by SPP5 and WG6, with the moderated $|E|^2$ amplitude and *Q*-factors (See Figure 4c).

## 3. Tuning resonant wavelength and *Q*-factors with high flexibility

Another strength of Al plasmonic-photonic crystal hybrid lie on the ability to tune its resonant wavelength with a high flexibility; this can be achieved simply by the change in the size of PS colloidal nanoparticles. As shown in Figure 5a, just varying the size of PS colloidal nanosphere, for example, from 230 nm to 330 nm, indeed enables the complement of the wide UV spectrum (from deep to far UV) of the achievable resonant SPP1 and WG2 modes, even though the other modes shows the limited range of the tunability, due to the scales law. The further extending the size range of PS colloidal nanosphere could allow other resonant modes to obtain wider UV spectrum, as with SPP1 and WG2.

Figure 5b summarizes the obtainable *Q*-factors at the varied resonant wavelengths, which were controlled by the adjustment of PS colloidal sizes; it is important to note that the *Q*-factor of WG1 mode can be dramatically increased to 1388 at a resonant wavelength of 270 nm (corresponding size of PS colloidal nanosphere is 230 nm). Also, the *Q*-factors of WG and SPP modes show a dispersive behavior, as shown in Figure 5b. In particular, even if some fluctuations are clearly visible, the *Q*-factors of WG modes are decreased, as the resonant wavelengths are increased. Meanwhile, the *Q*-factors of SPP modes are increased with the resonant wavelength. One of the key determinants for *Q*-factor is the intrinsic loss including the radiation losses resulting from the light coupling to the surrounding area and the ohmic metallic absorption losses [15, 18]. Given that the overall trend of *Q*-factor variations according to the resonant wavelengths can be directly associated with such losses in the system. In the case of WG modes, their resonant behaviors are supported almost in the dielectric PS nanospheres (see Figure 4c and its detailed modal analysis in Figure 6a-b). As these dielectric modes are maintained dominantly via total internal refraction, the refractive index contrast between PS colloidal monolayer and surrounding environments can have significant influence on the WG modes. Indeed, the lower refractive index of a flat Al film (0.2~0.4 at the UV wavelengths of interest, as shown in Figure 5c) than both that of air (1.0) and effective refractive index of the structure (1.38, average refractive index, calculated by weighting the filling ratio of PS colloidal monolayer in air) allows PS colloidal monolayer to minimize the radiative loss; greatly enhancing the *Q*-factors of WG modes. The trend of *Q*-factors evolution of WG modes according to the resonant wavelengths, which is well matched to the dispersive behavior of Al refractive index (Figure 5c), further supports this analysis.

In the case of SPP modes, as shown in Figure 4c and 7a-b (more detailed modal analysis of Figure 4c), both $E_x$ and $E_z$ can be highly confined within the vicinity between PS nanospheres and a flat Al film; thus, the ohmic absorptive loss of Al should have significantly influence on the *Q*-factors of SPP modes. Indeed, the *Q*-factors of SPP modes are much smaller than those of WG modes owing to large ohmic absorptive loss of Al; indeed, the dispersive lossy nature of Al, presented in the dispersion diagram of *k* of Al (Figure 5c), is well correlated with the trend of *Q*-factor enhancement with the resonant wavelengths.

## 4. Effect of Al$_2$O$_3$ on the resonant behaviors

We next demonstrate the effect of oxide contamination on the optical response of Al plasmonic-photonic crystals, as Al has been shown to be more vulnerable to oxide contamination, compared to Au [11]. These studies can make a contribution to setting out a general guide for a realistic experimental fabrication and reproducible plasmonic characterizations of Al plasmonic-photonic crystal hybrid. According to the recent result, reported by Naomi Halas group, two typical types of oxide (Al$_2$O$_3$) contaminations can preferably occur in the Al nanostructures: (i) Al$_2$O$_3$ encapsulation of the pristine Al nanostructure and (ii) uniform distribution of Al$_2$O$_3$ within Al (i.e., Al/Al$_2$O$_3$ composite) [11]. First, we investigated the effect of Al$_2$O$_3$ encapsulation layer on the resonant WG and SPP modes: herein, 300 nm-sized PS colloidal monolayers are assembled on the stacked Al$_2$O$_3$/Al/Si wafer (Figure 8a). The thickness of Al$_2$O$_3$ was varied from 1 nm to 5 nm, as this range of oxide layer thickness has been generally observed in an experimentally developed Al nanostructure [11]. In this case, the effect of Al$_2$O$_3$ encapsulation layer on the resonant WG and SPP modes can be interpreted by using perturbation theory [19]. According to perturbation theory, the material perturbation of the resonator can give rise to the shift of the resonant wavelengths, as follows:

$$\frac{\Delta \lambda_{res}}{\lambda_0} = \frac{-\iiint dV\left[(\Delta \vec{\mu} \cdot \vec{H}) \cdot \vec{H}_0^* + (\Delta \vec{\varepsilon} \cdot \vec{E}) \cdot \vec{E}_0^*\right]}{\iiint dV(\mu|\vec{H}_0|^2 + \varepsilon|\vec{E}_0|^2)}$$

where $\lambda_{res}$ and $\lambda_0$ indicate the perturbed and unperturbed resonant wavelength, respectively; both $\Delta \vec{\mu}$ and $\Delta \vec{\mu}$ represent the material perturbation by external sources; $\vec{H}_0$ and $\vec{E}_0$ present the unperturbed electromagnetic (EM) wave, whereas $\vec{H}$ and $\vec{E}$ correspond to the EM wave under perturbed status; the complex conjugates are expressed by $\vec{H}_0^*$ and $\vec{E}_0^*$. Thus, the dominator and numerator of the right side of the equation present the unperturbed total EM energy and the variation of EM energy by material perturbation, respectively. Conclusively, this theory indicates that the shift of the resonant wavelength can be effectively caused by the overlap between the material perturbation (herein, $Al_2O_3$ encapsulation layer) and the $|E|^2$.

Figure 8b summarizes the resonant wavelength shifts of WG and SPP modes (red shift), which are driven by the implementation of the $Al_2O_3$ encapsulation layer with different thickness. SPP5 mode becomes disappeared, when $Al_2O_3$ is inserted between Al and PS nanospheres; thus, we excluded this mode for clarity. From this result, we can observe that both SPP1 and SPP3 modes are more sensitive to the change in the thickness of $Al_2O_3$ encapsulation layer than WG2 and WG4 modes. This result is highly related with the relationship between $|E|^2$ spatial distribution and the material perturbation (i.e., $Al_2O_3$ encapsulation layer). Since both SPP1 and SPP3 modes can strongly confine the UV light between PS nanospheres and a flat Al film, we can expect that the overlap between materials perturbation ($Al_2O_3$ encapsulation layer) and $|E|^2$ can get maximized in the SPP1 and SPP3 modes. In contrast, for both WG2 and WG4 modes, most of $|E|^2$ is confined merely within the dielectric PS nanospheres, so as to induce a relatively small portion of the overlap between materials perturbation ($Al_2O_3$ encapsulation layer) and $|E|^2$. Indeed, the red shifts of these pure dielectric resonant wavelengths with respect to the change in the thickness of $Al_2O_3$ encapsulation layer is relatively limited, compared with those of SPP resonant modes. Interestingly, WG6 mode, despite of its main dielectric resonance, shows the red shifts of resonant wavelengths, comparable to the SPP1 and SPP3 modes, since certain portion of its resonant $|E|^2$ can be confined in the vicinity between PS nanospheres and Al layer as well as the inside of PS nanospheres (see Figure 4c).

Meanwhile, the obtainable $Q$-factors of WG and SPP modes together with the efficiency of light confinement also get reduced simultaneously, as the thickness of $Al_2O_3$ encapsulation layer increases (Figure 8c-d). Regarding WG2 and WG4 modes, the insertion of a relatively high refractive index $Al_2O_3$ (1.78-1.83 at the UV wavelengths) into the interface between PS nanospheres and Al can facilitate the radiative losses of such waveguided resonant modes; therefore, reducing the amplitude of light confinement and $Q$-factors, simultaneously. Regarding SPP1 and SPP3 modes, the $Al_2O_3$ encapsulation layer significantly hinders the intimate coupling between PS nanosphere and a metallic Al, in that the efficiency of the Al-mediated plasmonic light confinement and $Q$-factors are reduced (Figure 8c). These resonant wavelength red shifts and $Q$-factor changes appear to be quite sensitive to just few nanometer changes in the thickness of $Al_2O_3$ encapsulation layer; these exotic abilities can be used for the plasmonic/photonic reporter of oxide contamination.

Finally, we exploited the effect of another typed oxide contamination (i.e., $Al/Al_2O_3$ composite) on the optical response of Al plasmonic-photonic crystal hybrid: herein, $Al_2O_3$ encapsulation layer was set to be 0 nm in order to elucidate the effect of $Al/Al_2O_3$ composite. Toward this direction, there is an additional need to obtain the dielectric properties of $Al/Al_2O_3$ composite with varying the oxide ratio; the Bruggeman effective medium approximation allows us to calculate the complex permittivity of $Al/Al_2O_3$ composite, as follows [20]:

$$n_{Al}\frac{(\varepsilon_{Al}-\varepsilon)}{(\varepsilon_{Al}+2\varepsilon)}+n_{oxide}\left(\frac{\varepsilon_{oxide}-\varepsilon}{\varepsilon_{oxide}+2\varepsilon}\right)=0$$

where $n_{Al}$ and $n_{oxide}$ are the volume ratios of Al and $Al_2O_3$, respectively; $\varepsilon$ is the permittivity of $Al/Al_2O_3$ composite. The both $\varepsilon_{Al}$ and $\varepsilon_{oxide}$ were obtained by a modified Drude model and empirical measurement, respectively, as mentioned before. Figure 9a-c presents the calculated complex permittivity (Figure 9a-b) and refractive index (Figure 9c): with the ratio of $Al_2O_3$ to Al, the overall refractive index of $Al/Al_2O_3$ composite intrinsically increases.

Figure 10a shows the reflection spectra of 300 nm-sized PS colloidal monolayers onto the flat $Al/Al_2O_3$ composite with different oxide ratio. As the ratio of $Al_2O_3$ to Al increases, the intrinsic metallic properties of Al become weaken and get closed more to a dielectric material (see the change in the complex permittivity of $Al/Al_2O_3$ composite, according to the oxide ratio, as shown in Figure 9a-b). Therefore, eventually, three SPP modes (SPP1, SPP3, and SPP5) cannot be supported anymore, when the ratio of $Al_2O_3$ to Al reaches a relatively high ratio (40 %). Also, we can clearly see that two higher WG modes (i.e., WG4 and WG6) get emerged together in the end; only two WG modes (WG2 and WG6) are visible even at the high ratio of $Al_2O_3$ to Al (50%). In this case, the enhanced refractive index of back-plane substrate via adding 50 % of $Al_2O_3$ to pure Al results in the reduction of the refractive index contrast between PS colloidal monolayer and environments, so as to decrease both $Q$-factors and the light confinement, as shown in Figure 10b. Nonetheless, $Al/Al_2O_3$ composite with 50 % oxide can still show the lower refractive index (0.82 at the wavelength of interest) than both air (1.00) and PS colloidal monolayer (effective refractive index of 1.38); thus, can confine a certain portion of UV light within the dielectric PS nanosphere.

## 5. Conclusion

In this paper, we have demonstrated the efficient confinement of UV light into the Al plasmonic-photonic crystal hybrid, which can be developed in a highly versatile way: the self-assembly of PS colloidal monolayer onto a flat Al film. The stepwise modal analyses of the plasmonic/photonic crystal hybrid with respect to the surrounding environments (i.e., reflection spectra, $Q$-factors, and electric-field spatial distribution) details each resonant WG and SPP mode; highlighting the role of these resonant modes in the confinement of UV light within the structure. Importantly, we confirmed that Al plasmonic-photonic crystal hybrid can achieve the high $Q$-factors, at least one-order magnitude higher than other existing UV plasmonic nanostructures; thus, playing an important and versatile role in UV-mediated molecular sensing or spectroscopy. Finally, we highlighted the effect of the oxide contamination on the optical responses of Al plasmonic-photonic crystal hybrid, as this oxide contamination can be more realistic circumstance especially from the fabrication point of view. Thus, such collective set of theoretical exploitation can provide the general guide for the practical applications of Al plasmonic-photonic crystal hybrid to the efficient manipulation and confinement of UV light with a high flexibility.

**Acknowledgements**
This work was supported by Samsung Scholarship Research Fund 2014 program at Sungkyunkwan University (no. S-2014-0883-000).

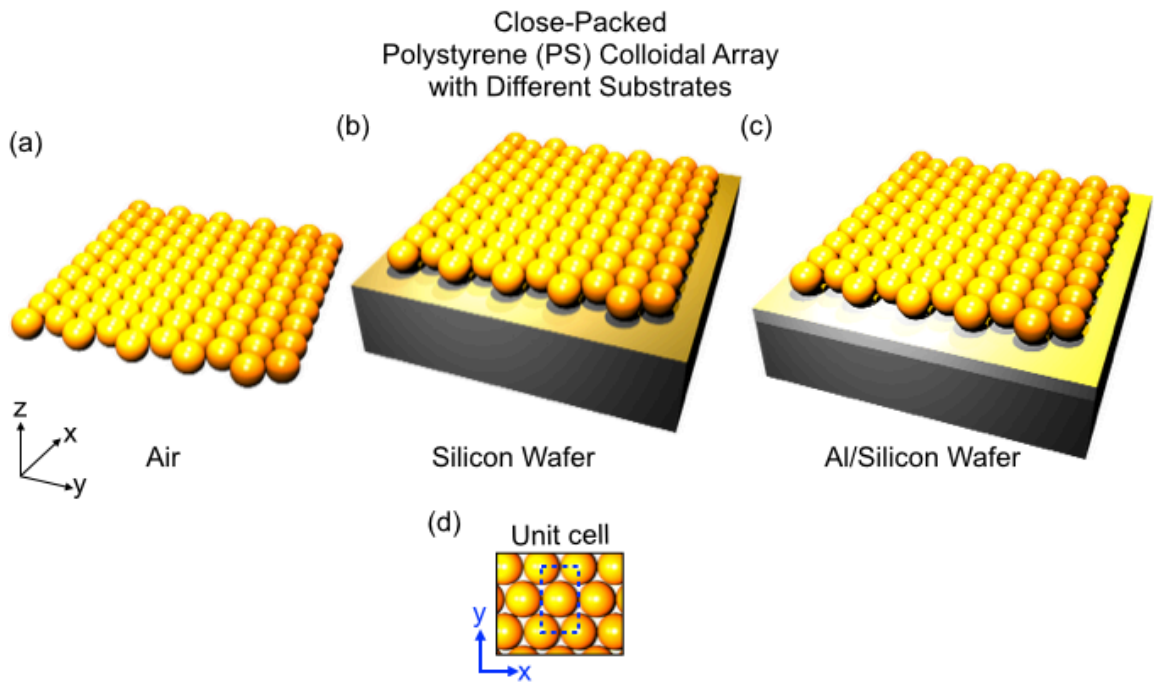

**Figure 1.** Schematic for three different structural primitives, numerically simulated in this paper: close-packed, polystyrene (PS) colloidal monolayer (a) in air; (b) onto silicon (Si) wafer; (c) onto 100 nm thick Al/Si wafer. (d) Schematic for unit cell of the model structure.

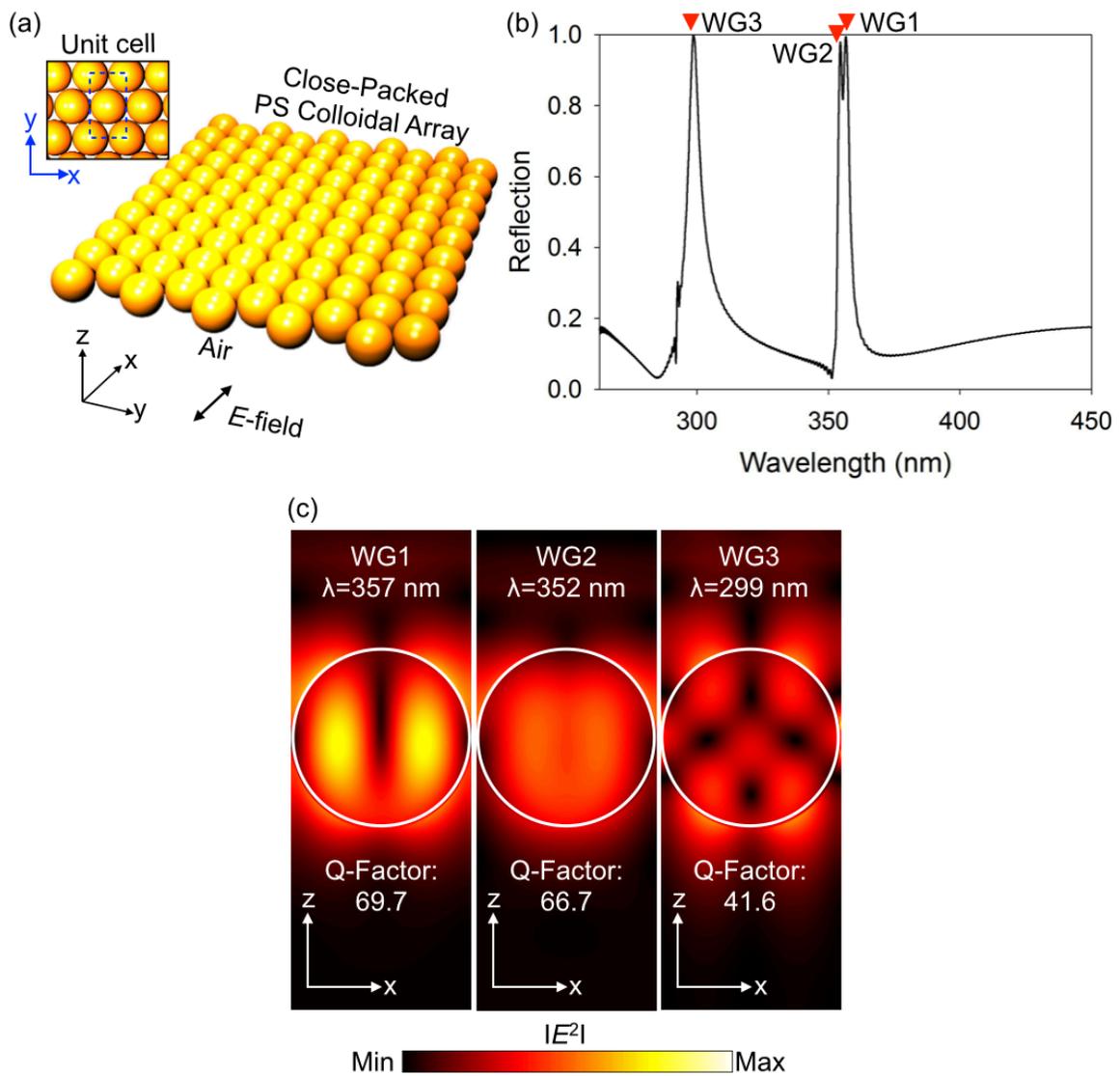

**Figure 2.** (a) Schematic for close-packed, 300 nm-sized PS colloidal monolayer in air. (b) UV regime reflection spectra of close-packed, 300 nm-sized PS colloidal monolayer in air. (c) Spatial distribution (*x-z* plane) of electric field intensity $|E|^2$ and *Q*-factors at WG1, WG2, and WG3 modes.

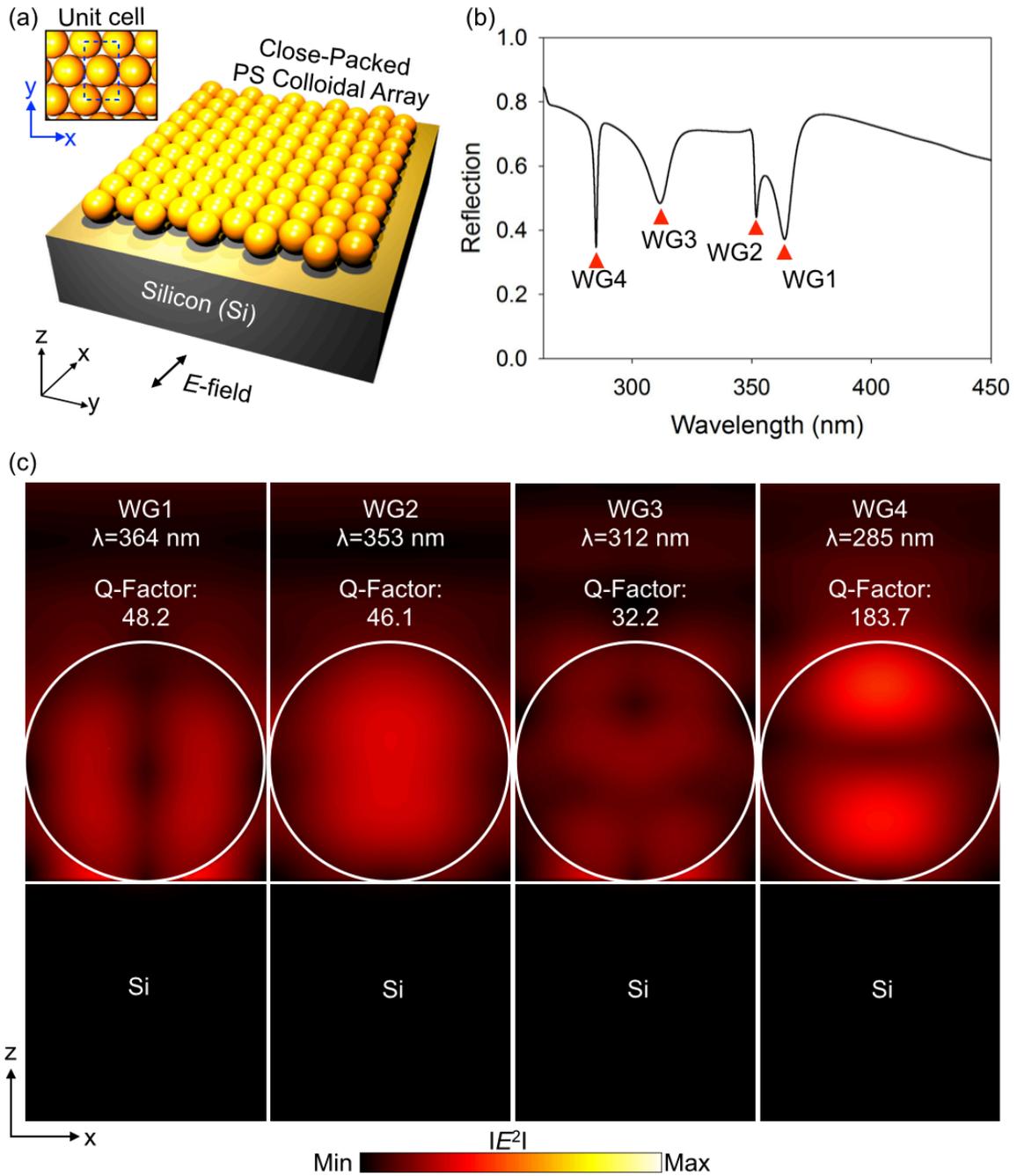

**Figure 3.** (a) Schematic for close-packed, 300 nm-sized PS colloidal monolayer, assembled onto Si wafer. (b) UV regime reflection spectra of close-packed, 300 nm-sized PS colloidal monolayer, assembled onto Si wafer. (c) Spatial distribution (*x-z* plane) of electric field intensity $|E|^2$ and *Q*-factors at WG1, WG2, WG3, and WG4 modes.

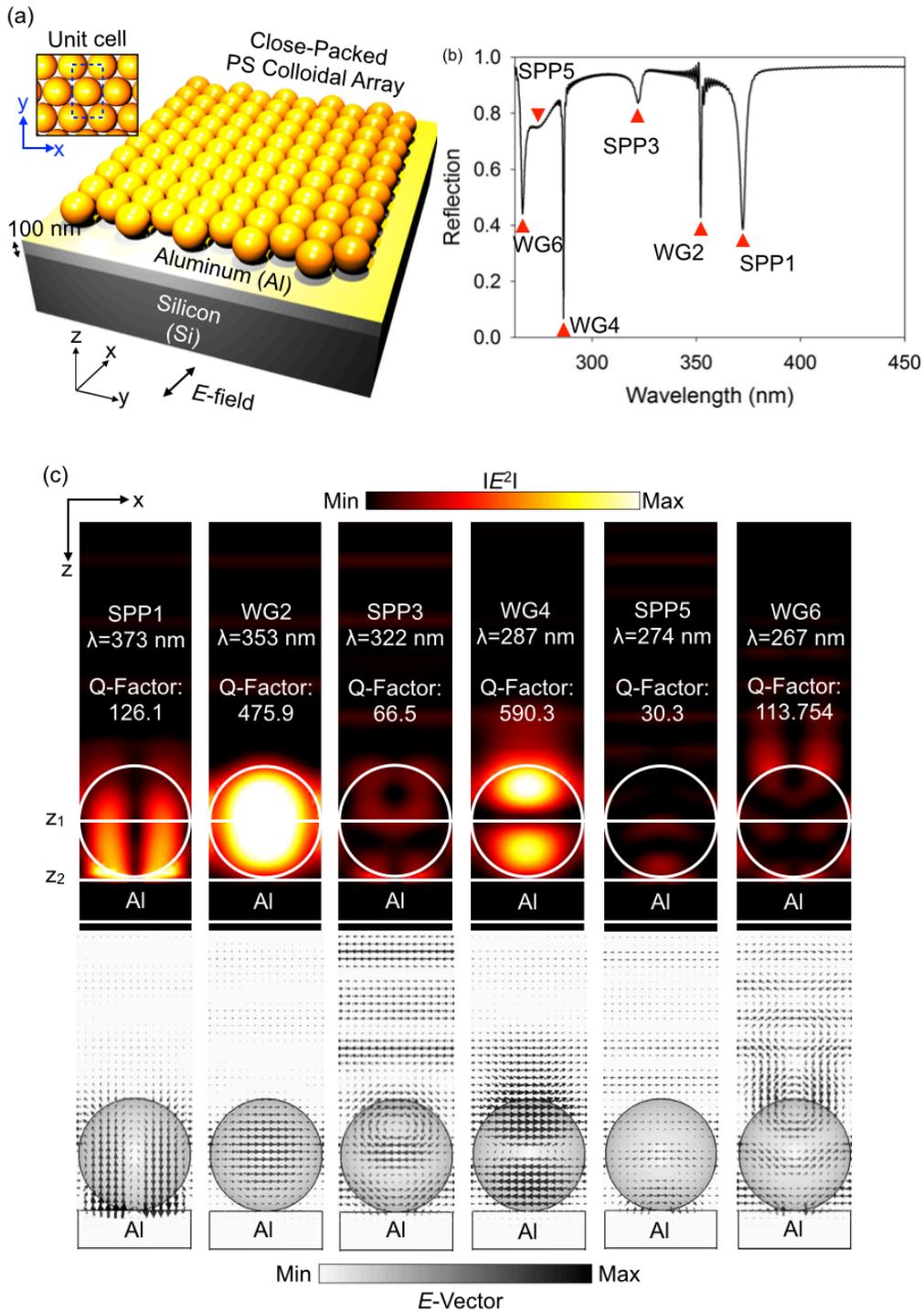

**Figure 4.** (a) Schematic for close-packed, 300 nm-sized PS colloidal monolayer, assembled onto 100 nm thick Al/Si wafer. (b) UV regime reflection spectra of close-packed, 300 nm-sized PS colloidal monolayer, assembled onto 100 nm thick Al/Si wafer. (c) Spatial distribution (*x-z* plane) of electric field intensity $|E|^2$/*Q*-factors (upper panel) and *E*-vector (bottom panel) at SPP1, WG2, SPP3, WG4, SPP5, and WG6 modes.

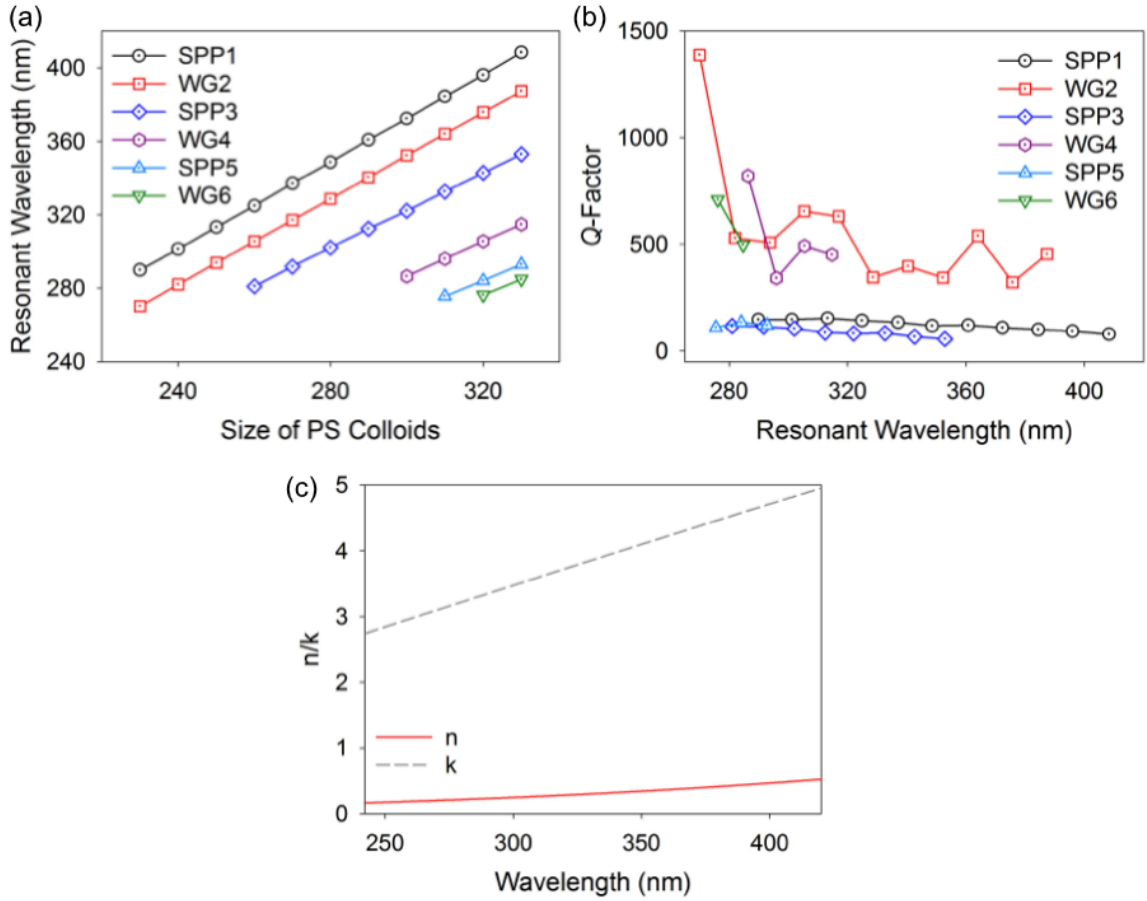

**Figure 5.** Tuning of (a) resonant wavelengthes and (b) *Q*-factors of each mode by the adjustment of the size of PS nanosphere, assembled onto a flat Al/Si wafer. (c) Dispersion diagram of Al refractive index.

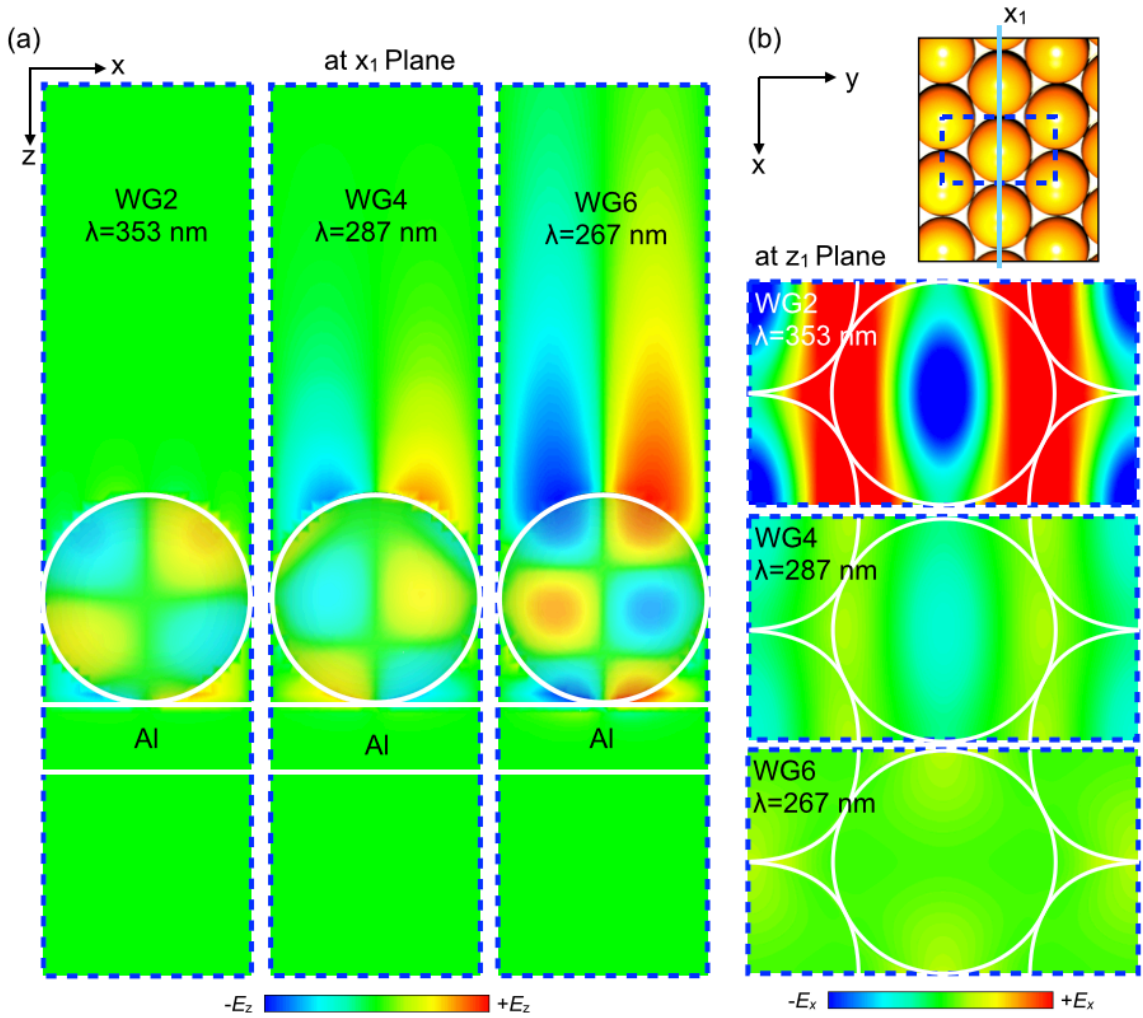

**Figure 6.** Resonant properties of WG2, WG4, and WG6 modes of close-packed, 300 nm-sized PS colloidal monolayer, assembled onto 100 nm thick Al/Si wafer. Spatial distribution of (a) $E_z$ at $x$-$z$ plane (at $x_1$ plane of schematic in Figure 6b) and (b) $E_x$ at $x$-$y$ plane (at $z_1$ plane of Figure 4c).

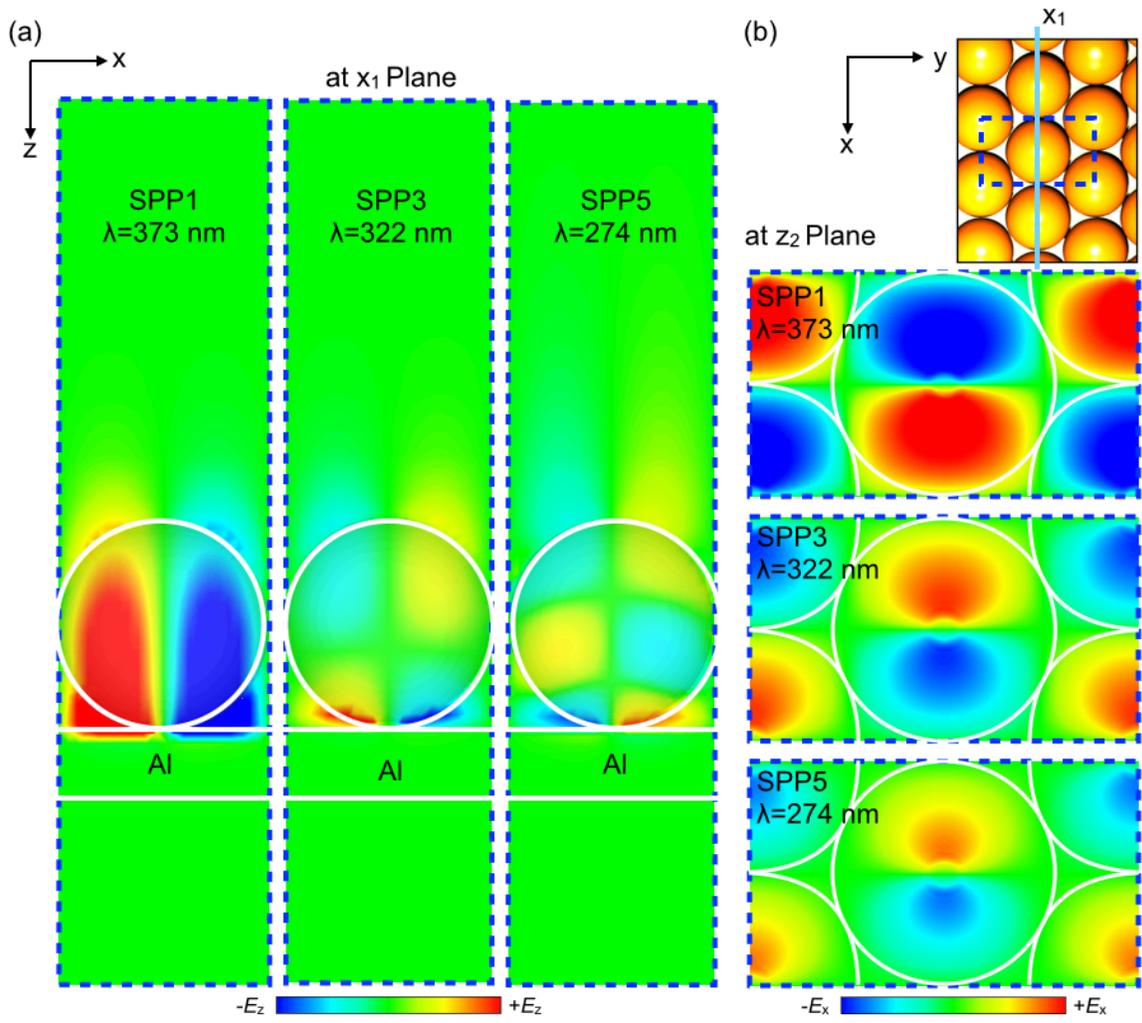

**Figure 7.** Resonant properties of SPP1, SPP3, and SPP5 modes of close-packed, 300 nm-sized PS colloidal monolayer, assembled onto 100 nm thick Al/Si wafer. Spatial distribution of (a) $E_z$ at $x$-$z$ plane (at $x_1$ plane of schematic in Figure 6b) and (b) $E_x$ at $x$-$y$ plane (at $z_2$ plane of Figure 4c).

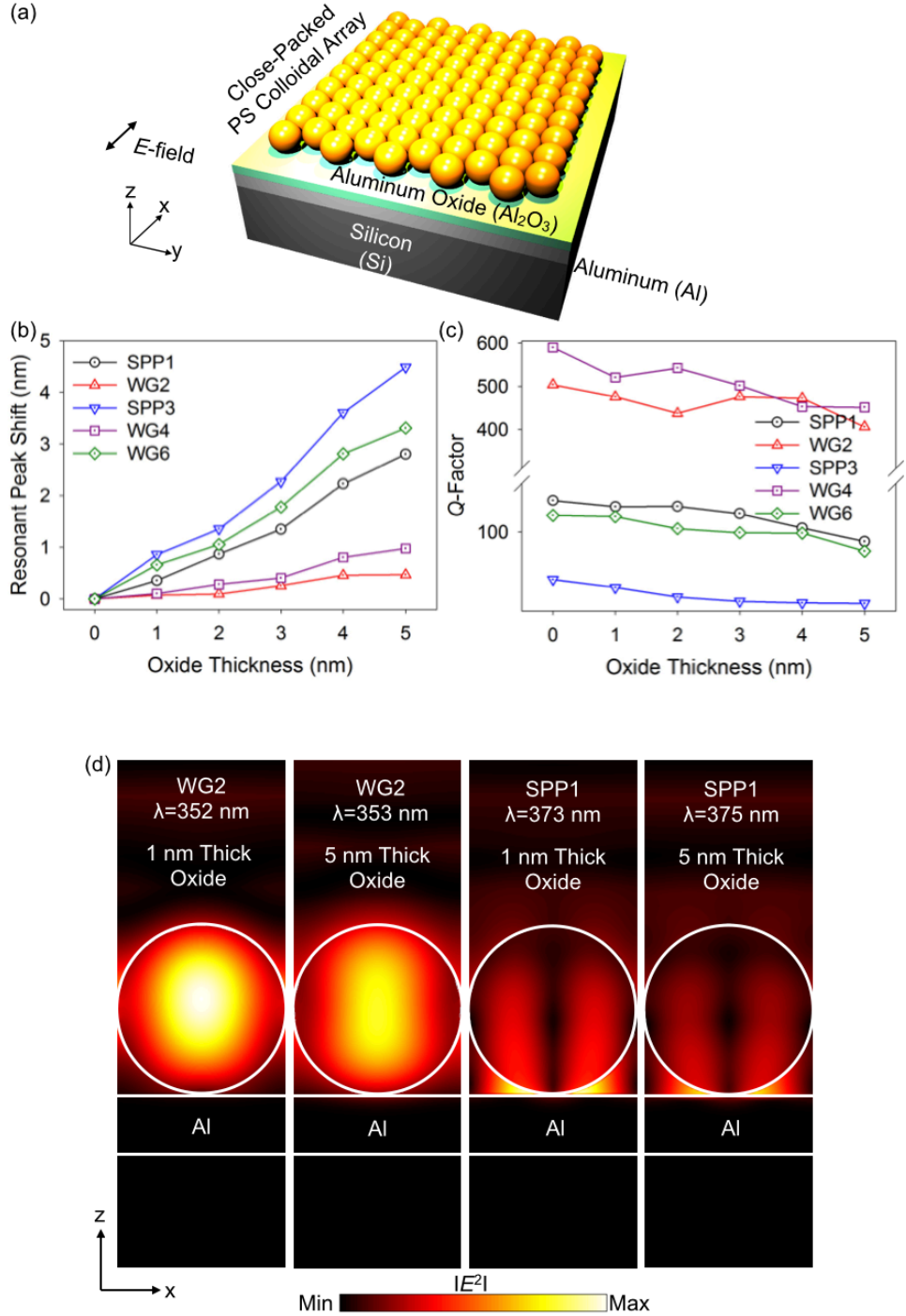

**Figure 8.** (a) Schematic for close-packed, 300 nm-sized PS colloidal monolayer, assembled onto $Al_2O_3$/Al/Si wafer. (b) The resonant wavelength shift of each modes according to the thickness of $Al_2O_3$ (from 1 nm to 5 nm). (c) $Q$-factors variations of each mode according to the thickness of $Al_2O_3$ from 1 nm to 5 nm. (d) Spatial distribution ($x$-$z$ plane) of electric field intensity $|E|^2$/$Q$-factors at WG2 and SPP1 modes, when the thickness of $Al_2O_3$ is 1 nm or 5 nm.

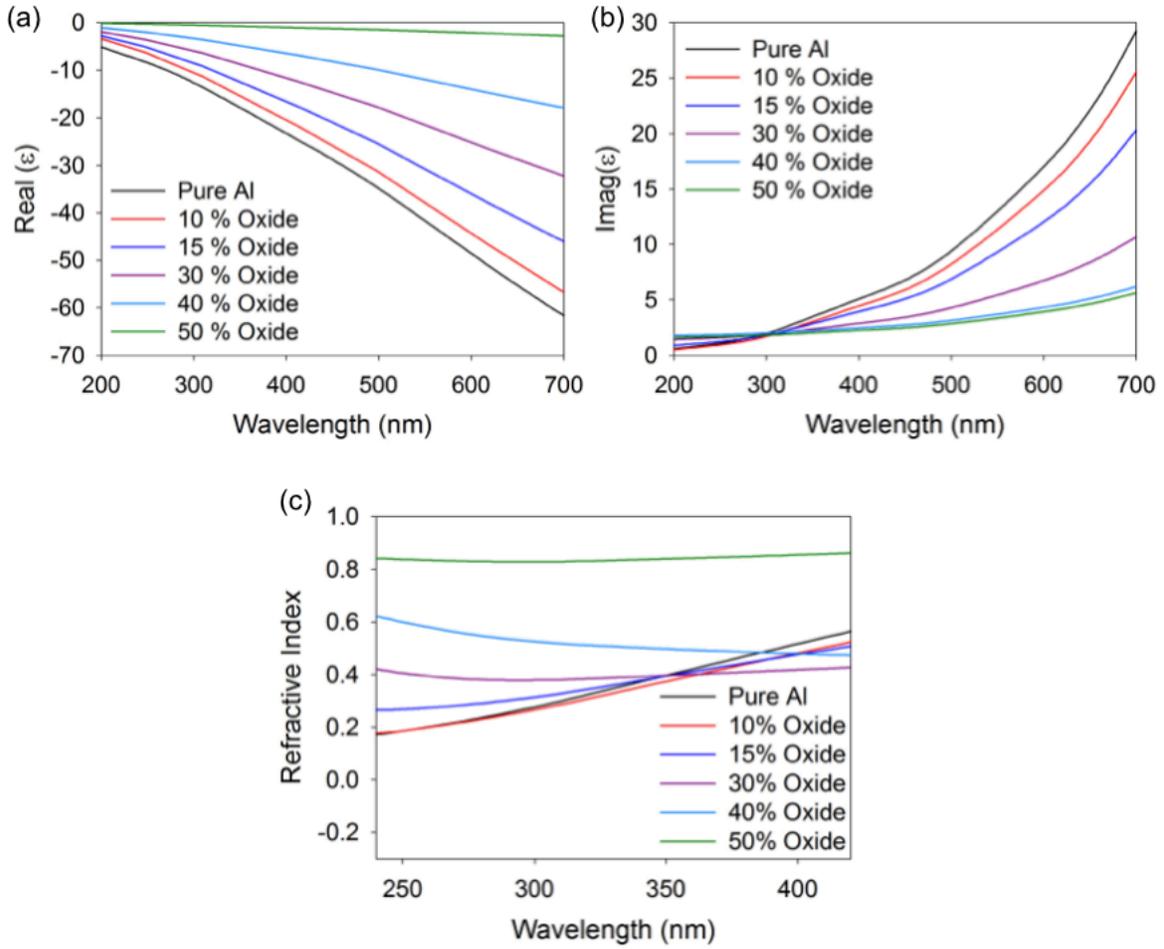

**Figure 9.** Complex permittivity of Al/Al$_2$O$_3$ composite with the controlled oxide ratio, which was calculated by Bruggeman model [20]. (a) Real permittivity. (b) Imaginary permittivity. (c) Refractive index.

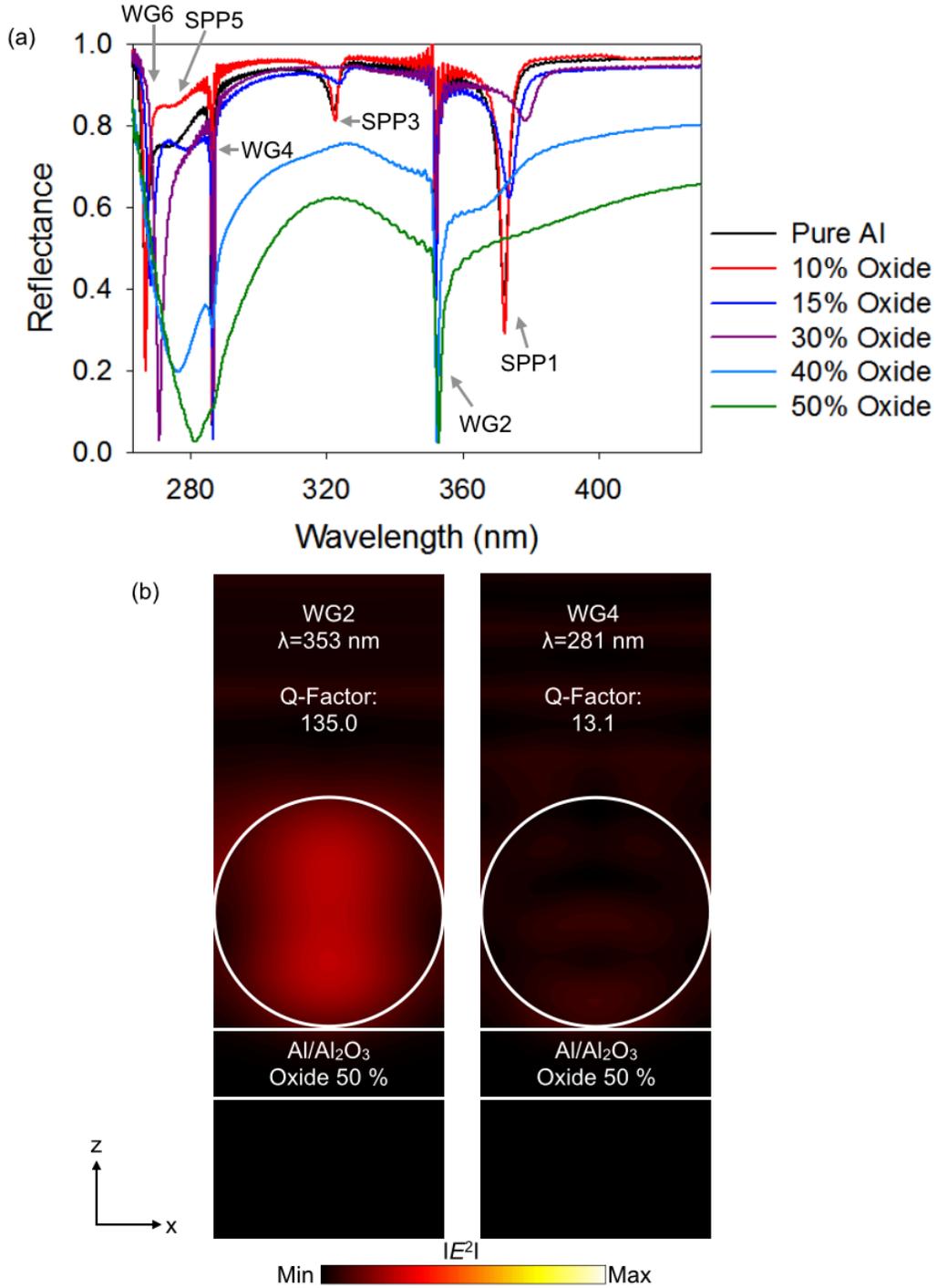

**Figure 10.** (a) UV regime reflection spectra of close-packed, 300 nm-sized PS colloidal monolayer, assembled onto 100 nm thick Al/Al$_2$O$_3$ composite with differernt oxide ratio. (b) Spatial distribution (*x-z* plane) of electric field intensity $|E|^2$ and *Q*-factors at WG2 and WG4 modes.